\documentclass[twocolumn,english,floats,showpacs,amsmath,amssymb,prb,aps]{revtex4}
\usepackage[T1]{fontenc}
\usepackage[latin9]{inputenc}
\usepackage{textcomp}
\usepackage{amstext}
\usepackage{graphicx}

\makeatletter

\providecommand{\tabularnewline}{\\}

\@ifundefined{textcolor}{}
{%
 \definecolor{BLACK}{gray}{0}
 \definecolor{WHITE}{gray}{1}
 \definecolor{RED}{rgb}{1,0,0}
 \definecolor{GREEN}{rgb}{0,1,0}
 \definecolor{BLUE}{rgb}{0,0,1}
 \definecolor{CYAN}{cmyk}{1,0,0,0}
 \definecolor{MAGENTA}{cmyk}{0,1,0,0}
 \definecolor{YELLOW}{cmyk}{0,0,1,0}
}

\usepackage{epsfig}

\makeatother

\usepackage{babel}
\begin{document}

\title{Anharmonic effects in the optical and acoustic bending modes of graphene}

\author{R. Ram\'{\i}rez, E. Chacón,}

\author{C. P. Herrero}

\affiliation{Instituto de Ciencia de Materiales de Madrid (ICMM), Consejo Superior
de Investigaciones Cient\'{\i}ficas (CSIC), Campus de Cantoblanco,
28049 Madrid, Spain }
\begin{abstract}
The out-of-plane fluctuations of carbon atoms in a graphene sheet
have been studied by means of classical molecular dynamic simulations
with an empirical force-field as a function of temperature. The Fourier
analysis of the out-of-plane fluctuations often applied to characterize
the acoustic bending mode of graphene is extended to the optical branch,
whose polarization vector is perpendicular to the graphene layer.
This observable is inaccessible in a continuous elastic model of graphene
but it is readily obtained by the atomistic treatment. Our results
suggest that the long-wavelength limit of the acoustic out-of-plane
fluctuations of a free layer without stress is qualitatively similar
to that predicted by a harmonic model under a tensile stress. This
conclusion is a consequence of the anharmonicity of both in-plane
and out-of-plane vibrational modes of the lattice. The most striking
anharmonic effect is the presence of a linear term, $\omega_{A}=v_{A}k$,
in the dispersion relation of the acoustic bending band of graphene
at long wavelengths ($k\rightarrow0$). This term implies a strong
reduction of the amplitude of out-of-plane oscillations in comparison
to a flexural mode with a $k^{2}$-dependence in the long-wavelength
limit. Our simulations show an increase of the sound velocity associated
to the bending mode, as well as an increase of its bending constant,
$\kappa,$ as the temperature increases. Moreover, the frequency of
the optical bending mode, $\omega_{O}(\Gamma$), also increases with
the temperature. Our results are in agreement with recent analytical
studies of the bending modes of graphene using either perturbation
theory or an adiabatic approximation in the framework of continuous
layer models. 
\end{abstract}


\maketitle

\section{Introduction\label{sec:intro}}

The crystalline order of a graphene layer has been focus of interesting
experimental investigations. Diffraction experiments by transmission
electron microscopy (TEM) reveal that suspended graphene sheets are
not perfectly flat: they exhibit intrinsic microscopic ripples. The
TEM atomic-resolution images display that the corrugations are static
with typical lengths in the range $L=$20-200 $\textrm{\AA}$ and
heights between $h=$2-20 $\textrm{\AA}$.\citep{meyer06} The bending
frequencies for wavelengths on the order of 200 $\textrm{\AA}$ are
estimated to be of order $10^{10}$ Hz (0.3 cm$^{-1}$). They are
fast for the time scale of electron diffraction experiments. Thus
the origin of the out-of-plane corrugation was suggested to be not
of thermal nature, but a consequence of adsorbed impurity atoms sitting
on random sites.\citep{thompson09} Nevertheless the exact origin
of the static corrugation in graphene is still unclear and probably
the stresses at the boundary of graphene during the device fabrication
play also an important role.\citealp{amorim16} The essential part
of anharmonicity in the corrugation behavior of graphene has been
stressed in a recent TEM study.\citep{kirilenko11} The root-mean-square
fluctuation of the graphene roughness was estimated as 1.7 $\textrm{\AA\ }$
at 300~K with a lateral scale of about 100 $\textrm{\AA}$. The most
striking result of this diffraction experiment, contrary to intuitive
expectation, was the \textit{increase} in the average corrugation
height with \textit{decreasing} temperature from 300~K to 150~K.\citep{kirilenko11} 

Atomistic simulations of the intrinsic ripples in graphene have predicted
that anharmonic couplings between bending and stretching modes significantly
diminish the mean-square height amplitude, $\left\langle h^{2}\right\rangle $,
of the out-of-plane thermal fluctuations. The relation of $\left\langle h^{2}\right\rangle $
to the number of atoms in the layer, $N$, has been described as a
power-law behavior, $N^{1-(\eta/2)}$, where $\eta$ is the roughness
exponent. The harmonic approximation for a typical flexural mode with
a quadratic dispersion relation predicts a vanishing exponent ($\eta$=0).\citep{gao14}
Note that this value represents a ``catastrophic'' divergence as
the mean-square height fluctuation grows as the area of the sheet,
$\left\langle h^{2}\right\rangle \propto N$. Consideration of anharmonic
effects by Nelson and Peliti results in a lower exponent $\eta$=1,
that diminishes the height fluctuations with respect to the harmonic
limit.\citep{nelson48} Interestingly, recent computer simulations
report anomalous exponents, $\eta$, that may vary depending on the
employed potential model and on the simulated ensemble (constant stress
or strain) in a range from $\eta$=0.67 to $\eta$=1.1.\citep{gao14,Los09}
It is believed that such anomalous exponents should be universal quantities,
therefore it remains unexplained the origin for the variability in
the roughness exponents reported in computer simulations.\citep{amorim16}

Analytical results from continuous models of graphene provide a picture
of the intrinsic surface corrugation that differs in some aspects
from the power-law behavior described by an anomalous roughness exponent.
The study of anharmonic effects by first-order perturbation theory
in Ref. \onlinecite{amorim14} shows that the dispersion relation
for the acoustic out-of-plane mode in graphene, $\omega_{A}(k)$,
is linear in the long-wavelength limit $(k\rightarrow0)$. The relation
$\omega_{A}=v_{A}k$, characteristic of sound waves at small $k$,
implies a roughness exponent $\eta$=2. This behavior is not related
to an external tension, i.e., the linear term has a finite value even
if the stress of the layer vanishes. An adiabatic approximation to
the anharmonic coupling between in-plane and out-of-plane acoustic
modes in graphene provides additional theoretical reasons to show
that the dispersion relation of the bending mode, $\omega_{A}(k)$,
must be necessary linear at small wavenumbers.\citealp{adamyan16}
It is remarkable that the perturbation analysis of Amorim \textit{et
al}.\citealp{amorim14} and the adiabatic approach of Adamayan \textit{et
al.},\citealp{adamyan16} even though they differ in the anharmonic
terms used to describe the phonon-phonon coupling, reach the same
conclusion. Namely the existence of a linear dispersion relation of
the bending mode at small wavenumbers. This term excludes, distinctively,
the appearance of power-law divergences in the mean-square amplitude,
$\left\langle h^{2}\right\rangle $, of the out-of-plane thermal fluctuations
as the area of the layer increases. The resulting amplitudes should
then display a much slower logarithmic grow as a function of the number
of atoms $N$.\citealp{gao14,adamyan16}

Here a series of classical molecular dynamics (MD) simulations of
a free suspended graphene sheet are presented using the empirical
long-range carbon bond order potential (LCBOPII).\citealp{los03,los05}
The focus lies on the characterization of the average height fluctuations
under conditions of zero stress and temperatures up to 2000~K. Finite
size effects have been considered by simulation of cells containing
between $10^{3}$ and $3\times10^{4}$ atoms. The analysis of the
simulations is based upon an atomistic model, which has the distinct
advantage over continuous models of providing information on both
the acoustic and optical oscillations in the direction perpendicular
to the layer. Particular emphasis is set upon the characterization
of effects related to the anharmonicity of the employed interatomic
potential. In this respect, an advantage of the numerical simulation
over analytical approaches is that the full anharmonicity of the potential
model is taken into account.

The structure of this paper is as follows. In Sec. \ref{sec:Fourier}
we summarize the Fourier analysis of the symmetric and antisymmetric
out-of-plane fluctuations. Basic equations are presented in Subsec.
\ref{sub:Basic-equations}, while a relation used to fit the $k-$dependence
of the acoustic height fluctuations is presented in Subsec. \ref{sub:omega_1D}.
The analysis of the simulation results is given in Sec. \ref{sec:results}.
The temperature dependence of the mean-square height fluctuations
of the acoustic modes is studied in Subsec. \ref{subamplitude_ZA},
while the related acoustic dispersion relation is the topic of Subsec.
\ref{sub:dispersion_ZA}. Anharmonic effects in the bending sound
velocity, bending rigidity and frequencies of optical out-of-plane
modes are studied in Subsec. \ref{sub:anharmonic-effects}. The divergence
of out-of-plane amplitudes with the system size is analyzed at 300~K 
in Subsec. \ref{sub:divergence_h2}. A brief discussion of the results
is presented in Sec. \ref{sec:Discussion}. Finally, we summarize
our conclusions in Sec. \ref{sec:conclusions}. Technical details
concerning the MD simulations are given in Appendix \ref{appendix:  MD}.

\section{Fourier analysis of out-of-plane fluctuations\label{sec:Fourier}}

MD simulations were performed in both $NVT$ and $NPT$ ensembles
($N$ being the number of atoms, $V$ is the area of the simulation
cell, $P$ the trace of the 2D stress tensor divided by 2, and $T$
the temperature). The simulation cell was defined by a supercell generated
with a two-dimensional (2D) rectangular cell, $(\mathbf{a},\mathbf{b})$.
For technical details concerning the simulation setup see Appendix
\ref{appendix:  MD}. Here we focus on the physics behind the Fourier
analysis of out-of-plane modes.

\subsection{Basic equations\label{sub:Basic-equations}}

The position of the \textit{j}'th atom of the simulation cell is represented
by a vector
\begin{equation}
\mathbf{r_{j}}=(\mathbf{u_{j}},z_{j})\;,
\end{equation}
where $\mathbf{u_{j}}$ is a 2D vector in the $(\mathbf{a},\mathbf{b})$
plane. The height of the atom is

\begin{equation}
h_{j}=z_{j}-\overline{z},
\end{equation}
with $\overline{z}=\sum_{j=1}^{N}z_{j}/N$ being the average height
of the layer. The carbon atoms in graphene are divided into two sublattices,
$\alpha$ and $\beta$, as shown in Appendix \ref{appendix:  MD}.
The discrete Fourier transform (dFT) of the heights of the carbon
atoms in the sublattice $\alpha$ is

\begin{equation}
H_{\alpha,n}=\frac{2}{N}\sum_{j=1}^{N/2}h_{j}e^{-i\mathbf{k_{n}u_{j}}}\;.\label{eq:fourier}
\end{equation}
Here the index $j$ runs only over $\alpha$ atoms. The set of $N_{k}$
vectors, $\mathbf{k_{n}}$, whose wavelengths are commensurate with
the simulation cell, is defined in Appendix \ref{appendix:  MD}.
A similar expression defines $H_{\beta,n}$ as the dFT of the heights
of the $\beta$ sublattice. One can define the dFT of the symmetric
and antisymmetric linear combinations of heights of $\alpha$ and
$\beta$ atoms

\begin{equation}
A_{n}=\frac{H_{\alpha,n}+H_{\beta,n}}{2}\;,\label{eq:acoustic}
\end{equation}

\begin{equation}
O_{n}=\frac{H_{\alpha,n}-H_{\beta,n}}{2}\;.\label{eq:optic}
\end{equation}
At the $\Gamma$ point, i.e., when $\mathbf{k_{n}}=\mathbf{0}$ in
Eq. (\ref{eq:fourier}), the phase difference between two atoms ($\alpha$
and $\beta$) in a hexagonal unit cell is 0 ($\pi$) for the symmetric
(antisymmetric) combination. However, for a generic $\mathbf{k_{n}}-$point
the phase difference is modulated by the value of the scalar product
$\mathbf{k_{n}u_{j}}$, that differs for $\alpha$ and $\beta$ atoms.
This phase modulation is similar to that encountered for the acoustic
and optical modes of a lattice with a base of two atoms.\citep{ashcroft}
We will see later that, for the $\mathbf{k_{n}}-$points within the
first hexagonal Brillouin zone (BZ), the module of the complex coefficients,
$\bar{A}_{n}$ and $\bar{O}_{n}$, are estimators for the amplitude
of the acoustic (ZA) and optical (ZO) vibrational modes of graphene
with polarization vector along the $z-$direction. 

The ensemble average height fluctuation
\begin{equation}
\left\langle h^{2}\right\rangle =\frac{1}{N}\left\langle \sum_{j=1}^{N}h_{j}^{2}\right\rangle \;,\label{eq:h2_def}
\end{equation}
is related to the set of spectral amplitudes $\left\langle \bar{A}_{n}^{2}\right\rangle $
and $\left\langle \bar{O}_{n}^{2}\right\rangle $ by the Parseval's
theorem of the dFT in Eq. (\ref{eq:fourier}). Taking into account
the definitions in Eqs. (\ref{eq:acoustic}) and (\ref{eq:optic}),
one gets

\begin{equation}
\left\langle h^{2}\right\rangle =\frac{2}{N}\left(\sum_{n=1}^{N_{k}}\left\langle \bar{A}_{n}^{2}\right\rangle +\left\langle \bar{O}_{n}^{2}\right\rangle \right)\;.\label{eq:parseval}
\end{equation}
Then, within an atomistic description of graphene, the average height
fluctuation is the sum of the symmetric and antisymmetric contributions.
We will quantify later the relative contribution of both modes.

Another interest of the the spectral amplitudes, $\left\langle \bar{A}_{n}^{2}\right\rangle $
or $\left\langle \bar{O}_{n}^{2}\right\rangle $, is their relationship
to the vibrational frequencies of the corresponding vibrational modes.
In the harmonic limit one has

\begin{equation}
\rho\omega_{A,n}^{2}=\frac{k_{B}T}{V_{a}\left\langle \bar{A}_{n}^{2}\right\rangle }\;,\label{eq:wn_An}
\end{equation}
where $k_{B}$ is the Boltzmann constant, $\rho=m/V_{a}$ is the atomic
density of the layer, $m$ is the atomic carbon mass, and $V_{a}=V/N$
the area per atom in the $x,y-$plane. Although this relation between
frequency and spatial amplitude is exact only in the harmonic limit,
it has been applied in the context of anharmonic vibrations of molecules
and solids as a reasonable linear response (LR) approximation.\citealp{ramirez01,ramirez05}
Anharmonic shifts in the stretching frequency of hydrogen molecules
adsorbed as isolated impurities in graphite and silicon were studied
by this method.\citealp{herrero09,herrero10} Within this LR approximation
anharmonic vibrational frequencies are estimated with Eq. (\ref{eq:wn_An})
from\textit{ anharmonic vibrational amplitudes} that are obtained
by computer simulations. 

\begin{figure}
\vspace{-1.7cm}
~\hspace{-0.4cm}
\includegraphics[width= 8.5cm]{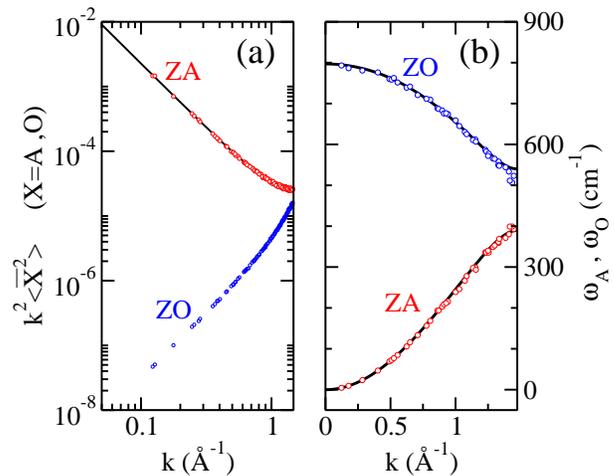}
\vspace{-0.5cm}
\caption{(a) Log-log plot of the spectral amplitudes of the symmetric (acoustic,
ZA) and antisymmetric (optical, ZO) modes of graphene derived from
$NPT$ simulations at 1~K. The $\mathbf{k_{n}}-$points, defined in
Eq. (\ref{eq:k_n}), correspond to a simulation cell with 960 atoms.
The symbols show the square of the amplitude times $k^{2}$ as a function
of module of the wavevector $k$. The continuous line is the least
squares fit of the acoustic branch to Eq. (\ref{eq:fit_k2_A}). The
largest displayed $k-$point corresponds to the point $M$ at the
boundary of the hexagonal BZ ($k_{M}=1.48\;\textrm{\AA}^{-1}$). (b)
The dispersion relations of the ZA and ZO bands of graphene, as derived
by Eq. (\ref{eq:wn_An}), are shown by circles. The continuous lines
are calculated by diagonalizing the dynamical matrix of the LCBOPII
model along the $\Gamma M$ direction of the hexagonal BZ. }
\label{fig:1K}
\end{figure}

As illustration of the physical information of the spectral functions
$A_{n}$ and $O_{n}$, we have derived them in a classical $NPT$
simulation of graphene with $N=960$ atoms at $P=0$ and $T=1$~K.
This temperature is chosen deliberately low with the purpose of having
vibrational modes close to their harmonic limit. The simulation data
can be then checked against analytical results. 

In Fig. \ref{fig:1K}a the ensemble average of the dimensionless quantities
$k_{n}^{2}\left\langle \bar{A}_{n}^{2}\right\rangle $ and $k_{n}^{2}\left\langle \bar{O}_{n}^{2}\right\rangle $
is shown as a function of the module of the wavevector, $k=\mathbf{\left|k\right|}$.
The graphical representation as a function of $k$ (instead of the
vector $\mathbf{k}$) is justified by the in-plane isotropy of graphene.
The isotropy is particularly valid in the elastic long-wavelength
limit ($k\rightarrow0)$, although less true when the vector $\mathbf{k}$
approaches the boundary of the 2D hexagonal BZ.\citealp{lambin14}
As expected, the amplitudes for the symmetric branch are always larger
that those of the asymmetric branch, and the difference increases
in the long-wavelength limit. 

The mean-square height $\left\langle h^{2}\right\rangle $ obtained
in the $NPT$ simulation at 1~K is $\left\langle h^{2}\right\rangle =5.9\times10^{-4}$
$\textrm{\AA}^{2}$. The contribution from the symmetric and antisymmetric
modes derived by Eq. (\ref{eq:parseval}) amounts to 89\% and 11\%,
respectively. Note that the contribution of the antisymmetric mode
to the height fluctuation $\left\langle h^{2}\right\rangle $ is significant.
However to the best of our knowledge this contribution has never been
quantified in previous simulations of graphene.\citealp{gao14,roldan11,Los09} 

The wavenumbers, $\omega_{A}$ and $\omega_{O}$, derived by Eq. (\ref{eq:wn_An})
from the amplitudes of the ZA and ZO modes are displayed as circles
in Fig. \ref{fig:1K}b. For comparison, the continuous lines show
the frequencies obtained by diagonalizing the dynamical matrix of
graphene along the $\Gamma-M$ direction of the hexagonal BZ. The
dynamical matrix was calculated with the same potential model (LCBOPII)
as employed in the simulations. Vibrational frequencies of \textit{both}
acoustic and optical branches are reproduced accurately by the analysis
of spectral amplitudes. It is remarkable that one gets realistic vibrational
frequencies even near the boundary of the first hexagonal BZ. This
one-to-one correspondence between symmetric (antisymmetric) out-of-plane
fluctuations and acoustic (optical) vibrational amplitudes is somewhat
lost when the vector $\mathbf{k_{n}}$ lies outside the first BZ.
The spatial relation between the $\mathbf{k_{n}}-$grid and the hexagonal
BZ is displayed in Fig. \ref{fig:k} in Appendix \ref{appendix:  MD}.
The relative large contribution (11 \%) of the asymmetric band to
$\left\langle h^{2}\right\rangle $ is caused by the increasing acoustic
character of the asymmetric out-of-plane fluctuations at $k$ values
larger than those shown in Fig. \ref{fig:1K}. 

\begin{table*}
\caption{Parameters $D,L$, and $C$ obtained from Eq. (\ref{eq:fit_k2_A})
by least squares fits of the simulated values of $k_{n}^{2}\left\langle \bar{A}_{n}^{2}\right\rangle $
at several temperatures. The fits were performed in the $k-$interval
defined by $k<1$ $\textrm{\AA}^{-1}$. $V_{a}$ is the area per atom.
The last columns are the linear term of the acoustic dispersion relation,
$\sigma$, and the bending rigidity, $\kappa$. The results correspond
to a simulation cell with 960 atoms.}
\centering{}\label{tab:1}%
\begin{tabular}{lllccccc}
\hline
$T$ (K) &  & $D$ (eV$\textrm{\AA}^{-4}$) & $L$ ($\textrm{\AA}$) & $\quad$$C$$\quad$ & $V_{a}$ ($\textrm{\AA}^{2}$/atom) & $\sigma$ (eV$\textrm{\AA}^{-2}$) & $\kappa$ (eV)\tabularnewline
\hline
1 &  & 4.716 & \multicolumn{1}{c}{1.491} & 0.2500 & 2.6189 & 0.000 & 1.49\tabularnewline
50 &  & 3.320 & 1.645 & 0.2499 & 2.6185 & 0.001 & 1.52\tabularnewline
300 &  & 2.904 & 1.727 & 0.2491 & 2.6173 & 0.008 & 1.61\tabularnewline
1000 &  & 1.693 & 2.059 & 0.2480 & 2.6183 & 0.014 & 1.88\tabularnewline
2000 &  & 1.292 & 2.280 & 0.2474 & 2.6279 & 0.018 & 2.15\tabularnewline
\hline
\end{tabular}
\end{table*}

The realistic prediction of the ZA and ZO vibrational bands in Fig.
\ref{fig:1K}b encourages us to apply this spectral analysis at higher
temperatures, where anharmonic effects are expected to be relevant.
However, an additional numerical tool would be helpful for the study
of the long-wavelength limit of the acoustic modes. Namely a realistic
analytical function to fit the $k-$dependence of its spectral amplitude.

\subsection{Atomistic model for the acoustic spectral amplitudes\label{sub:omega_1D}}

The phenomenological dispersion relation for the acoustic branch of
a continuous membrane is 

\begin{equation}
\rho\omega_{A}^{2}=\sigma k\text{\texttwosuperior\ }+\kappa k^{4}\;,\label{eq: w=00003Dk2_k4}
\end{equation}
where $\sigma$ is the external stress, and $\kappa$ is the bending
rigidity. This relation could be used, with the help of Eq. (\ref{eq:wn_An}),
to fit the $k-$dependence of the function $k_{n}^{2}\left\langle \bar{A}_{n}^{2}\right\rangle $
(see Fig. \ref{fig:1K}a). However the interval $\left[0,k_{A}\right]$,
where the phenomenological expression is valid, is not clearly defined.
Therefore it is convenient to work with an improved dispersion relation
for graphene based on an atomistic model instead of a continuous limit
as in Eq. (\ref{eq: w=00003Dk2_k4}). 

The simplest atomic model that displays an acoustic flexural mode
is a one-dimensional chain of atoms with interactions up to second
nearest neighbors. The dispersion relation for this model has the
following analytical form (see Appendix \ref{appendix:w})

\begin{equation}
\rho\omega_{A}^{2}=D\left[\sin^{2}\left(Lk/2\right)-C\sin^{2}\left(Lk\right)\right]\;,\label{eq:w_sin_2_DLC}
\end{equation}
where $D,L$ and $C$ are treated here as adjustable parameters. The
Taylor expansion of this analytical function contains only even powers
of $k.$ The first two coefficients, as defined in Eq. (\ref{eq: w=00003Dk2_k4}),
are

\begin{equation}
\sigma=DL^{2}\left(\frac{1}{4}-C\right)\;,\label{eq:sigma}
\end{equation}

\begin{equation}
\kappa=DL^{4}\left(\frac{C}{3}-\frac{1}{48}\right)\;.\label{eq:kappa}
\end{equation}
Following Eqs. (\ref{eq:wn_An}) and (\ref{eq:w_sin_2_DLC}), the
simulated results of $k_{n}^{2}\left\langle \bar{A}_{n}^{2}\right\rangle $
will be fitted by a least squares method to the function

\begin{equation}
f(k)=\frac{k_{B}T}{V_{a}}\frac{k^{2}}{D\left[\sin^{2}\left(Lk/2\right)-C\sin^{2}\left(Lk\right)\right]}\;,\label{eq:fit_k2_A}
\end{equation}
that depends on the parameters $D,L$ and $C$. All the fits in this
work were performed with $k-$points satisfying $k<1\;\textrm{\AA}^{-1}$.

The continuous line in Fig. \ref{fig:1K}a shows the fit of $k_{n}^{2}\left\langle \bar{A}_{n}^{2}\right\rangle $
for the simulation at 1~K. The fitted parameters are summarized in
the first line of Tab. \ref{tab:1}. The value of the parameter $C=1/4$
implies that $\sigma=0$ here. A value of $\kappa=1.49$ eV is derived
from Eq. (\ref{eq:kappa}). We have checked that this value agrees,
within the statistical error, with the numerical second derivative
of $\omega_{A}$ 
\begin{equation}
\left(\frac{\kappa}{\rho}\right)^{1/2}=\frac{1}{2}\left(\frac{\partial^{2}\omega_{A}}{\partial k^{2}}\right)_{k=0}\;.
\end{equation}
Here $\omega_{A}$ was calculated by diagonalizing the dynamical matrix
of the employed LCBOPII model. This $\omega_{A}$ band was shown by
a continuous line in Fig. \ref{fig:1K}b. The numerical second derivative
at $k=0$ gives the value $\kappa=1.48$ eV.

\section{simulation results\label{sec:results}}

$NPT$ simulations of graphene were performed in the classical limit
at external stress $P=0.$ Two cell sizes were employed to study temperatures
in the range 50-2000~K. Predictions based on simulations with $N=960$
atoms were checked against the results obtained with a larger cell
with $8400$ atoms. 

\begin{figure}
\vspace{-0.6cm}
\includegraphics[width= 8cm]{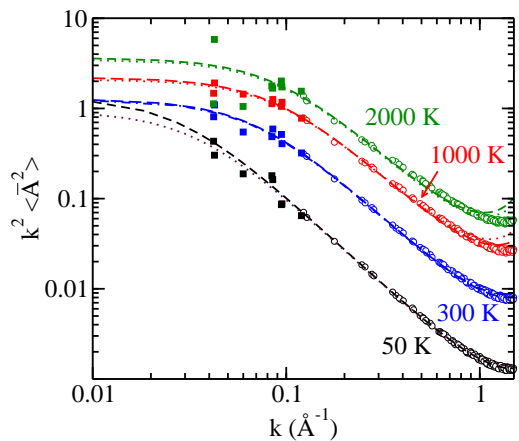}
\vspace{-0.0cm}
\caption{Log-log plot of the spectral amplitudes of the ZA modes of graphene
derived from $NPT$ simulations at several temperatures. Open circles
are results with $N=960$ atoms for $k>0.12$ $\textrm{\AA}^{-1}$.
Closed squares correspond to $N=8400$ for $0.04<k<0.12$ $\textrm{\AA}^{-1}$.
The broken and dotted lines are least squares fits of the simulation
results to Eq. (\ref{eq:fit_k2_A}). Broken lines are for $N=960$
atoms and dotted ones for $N=8400$. The broken and dotted lines at
the same temperature are almost indistinguishable except at 50~K in
the region of low $k$.}
\label{fig:A_t}
\end{figure}

\subsection{Spectral amplitudes for ZA modes\label{subamplitude_ZA}}

The values of $k_{n}^{2}\left\langle \bar{A}_{n}^{2}\right\rangle $
for $N=960$ are displayed as open circles in Fig. \ref{fig:A_t}.
The studied temperatures are 50, 300, 1000, and 2000~K. The size of
the simulation cell implies that the shortest wavevector for the out-of-plane
oscillations has $k=0.12$ $\textrm{\AA}^{-1}$. The largest displayed
$k$ corresponds to the point $M$ at the boundary of the hexagonal
BZ. Least squares fits of the simulation data by Eq. (\ref{eq:fit_k2_A})
are shown by broken lines. The fitted coefficients $D,L,$ and $C$
are summarized in Tab. \ref{tab:1}. The fitted functions follow accurately
the simulation data in the displayed $k-$interval. Only at high temperature
($T\geq$1000~K) there appears a small deviation between the fitted
function and simulation data for $k>1$ $\textrm{\AA}^{-1}$. 

In the region with $k<0.12$ $\textrm{\AA}^{-1}$, i.e. for long-wavelength
oscillations, the fitted functions represent obviously an extrapolation
of the simulation data. The extrapolation clearly predicts a flattening
of the function $k^{2}\left\langle \bar{A}^{2}\right\rangle $ at
the four studied temperatures. This flattening is absent in the harmonic
limit displayed in Fig. $\ref{fig:1K}$a. Numerically, $C$ is the
parameter that controls the behavior of the function $k^{2}\left\langle \bar{A}^{2}\right\rangle $
at low $k$. If the coefficient $C$ becomes smaller than 1/4 {[}see
Eq. (\ref{eq:sigma}){]}, then the dispersion relation of the ZA modes
displays a linear term, $\sigma>0$. The fitted $C$ coefficients
in Tab. \ref{tab:1} decrease as the temperature increases. The value
$\sigma>0$ predicted by the simulations \textit{at zero stress} is
an anharmonic effect activated by the temperature. 

As a consistency check for the extrapolated behavior of $k^{2}\left\langle \bar{A}^{2}\right\rangle $
at low $k$, the corresponding values for a larger cell with 8400
atoms are represented as closed squares in Fig. \ref{fig:A_t}. To
avoid an overcrowding of points only those wavevectors with $k<0.12$
$\textrm{\AA}^{-1}$ are plotted. The displayed squares correspond
to oscillations with wavelengths ($\lambda=2\pi/k$) that are inaccessible
to the simulations with 960 atoms. The new points in the region $0.04<k<0.12$
$\textrm{\AA}^{-1}$ follow with reasonable accuracy the functions
fitted with the smaller cell. This is true for the four studied temperatures.
The simulation results of $k_{n}^{2}\left\langle \bar{A}_{n}^{2}\right\rangle $
with 8400 atoms have been also fitted with Eq. (\ref{eq:fit_k2_A}).
The functions are plotted as dotted lines in Fig. \ref{fig:A_t}.
The dotted lines are nearly indistinguishable from the fits with the
smaller cell (broken lines). A small difference at $k<0.04$ $\textrm{\AA}^{-1}$
appears only at 50~K. We consider this agreement as a strong evidence
that the dispersion relation in Eq. (\ref{eq:w_sin_2_DLC}), which
is the basic ingredient for the fit of the spectral amplitudes $k_{n}^{2}\left\langle \bar{A}_{n}^{2}\right\rangle $,
provides a physically sound atomistic approximation for the out-of-plane
acoustic oscillations of graphene. 

\begin{figure}
\vspace{-1.0cm}
~\hspace{-0.2cm}
\includegraphics[width=8.5cm]{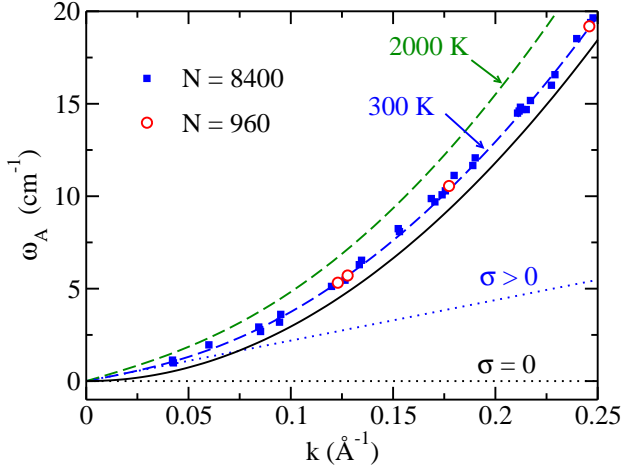}
\caption{Temperature dependence of the dispersion band of the ZA modes in the
long-wavelength region. The continuous line is the classical $T\rightarrow0$
limit, derived by diagonalizing the dynamic matrix along the $\Gamma M$
direction. The broken lines corresponds to Eq. (\ref{eq:w_sin_2_DLC})
with the parameters $D,L,$ and $C$ obtained from the fits of $k_{n}^{2}\left\langle \bar{A}_{n}^{2}\right\rangle $
with 960 atoms at 300 and 2000~K, respectively (see Tab \ref{tab:1}).
The open circles are the results derived from Eq. (\ref{eq:wn_An})
with $960$ atoms at 300~K . The closed squares are the corresponding
results for $8400$ atoms at 300~K. The straight dotted line with
positive slope $(\sigma>0)$ is the linear term of the dispersion
curve at 300~K. The slope is zero $(\sigma=0)$ for the quadratic
dispersion in the limit $T\rightarrow0$.}
\label{fig:w_za}
\end{figure}

\subsection{Dispersion relation of ZA modes\label{sub:dispersion_ZA}}

The fitted constants $D,L$ and $C$ in Tab. \ref{tab:1} allow us
to plot the acoustic dispersion relation, $\omega_{A}(k)$, according
to Eq. (\ref{eq:w_sin_2_DLC}). The curves at 300~K and 2000~K are
plotted as broken lines in Fig. \ref{fig:w_za}. The dispersion relations
are shown up to $k=0.25$ $\textrm{\AA}^{-1}.$ For reference we also
display the harmonic $T\rightarrow0$ limit derived by diagonalization
of the dynamical matrix of the employed LCBOPII model, which was already
plotted in Fig. \ref{fig:1K}b. This limit, shown by a full line,
displays a quadratic dispersion with vanishing linear term ($\sigma=0$)
as $k\rightarrow0$. However, the dispersion curves at 300 and 2000~K 
show finite linear terms ($\sigma>0$) as $k\rightarrow0$. The
dotted line with a positive slope displays the linear term at 300~K.

In Fig. \ref{fig:w_za} we have also plotted the discrete frequencies,
$\omega_{A,n}$, derived from the spatial amplitudes $\left\langle \bar{A}_{n}^{2}\right\rangle $
by Eq. (\ref{eq:wn_An}). The open circles are frequencies obtained
from the simulation with 960 atoms. As expected, the open circles
are in good agreement with the broken curve at 300~K, because both
data were evaluated from the same set of $\left\langle \bar{A}_{n}^{2}\right\rangle $
values. However, the set of discrete frequencies derived from the
simulation with 8400 atoms at 300~K provide an independent check of
the results obtained with 960 atoms. These frequencies are plotted
as closed squares in Fig. \ref{fig:w_za}. The density of sampled
$k-$points is much larger than for 960 atoms. The closed squares
are in reasonable agreement to the broken line predicted by the smaller
cell at 300~K. It is remarkable that the dispersion relation, $\omega_{A}(k)$,
displays a very small size effect, in the sense that a simulation
with only 960 atoms seems to provide a reasonably converged result
for this function.

The deviation of the dispersion curve at finite temperature from the
harmonic $T\rightarrow0$ limit is an anharmonic effect predicted
by the simulation. Note that at the lowest $k$ accessible in our
simulations ($k=0.04\;\textrm{\AA}^{-1},\;\lambda=150$ $\textrm{\AA}$)
the dispersion curve is very close to the straight line that plots
its linear term. The estimated frequency for this $k$ is only 1 cm$^{-1}$
($3\times10^{11}$ Hz) at 300~K, about two times larger than the harmonic
$T\rightarrow0$ limit. This anharmonic shift of the $\omega_{A}$
frequency is small in absolute value, but has a large effect for the
out-of-plane carbon fluctuations. A related important anharmonic effect
is that the sound velocity of the ZA branch, defined as

\begin{equation}
v_{A}=\left(\frac{\partial\omega_{A}}{\partial k}\right)_{k=0}=\left(\frac{\sigma}{\rho}\right)^{1/2}\;,
\end{equation}
becomes finite. At 300~K the sound velocity amounts to 0.4 km/s, while
at 2000~K increases to 0.6 km/s. Our results are lower than the numerical
estimations based on the adiabatic model of Adamyan \textit{et al.}
which report a value of 1.1 km/s at 2000~K.\citealp{adamyan16} 

In the following Subsection we quantify the anharmonic effects of several
important magnitudes related to the out-of-plane carbon fluctuations. 

\begin{figure}
\vspace{-1.3cm}
\includegraphics[width= 9cm]{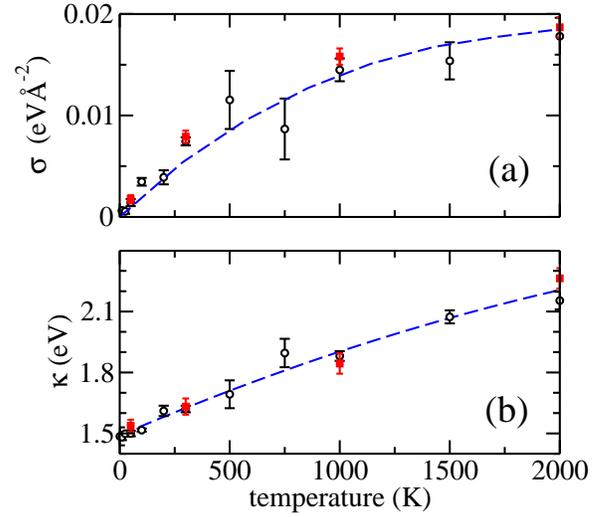}
\vspace{-0.5cm}
\caption{(a) Temperature dependence of the linear coefficient, $\sigma$, of
the dispersion relation of the ZA mode as derived from MD simulations
up to 2000~K. (b) Bending rigidity $\kappa$ of graphene as a function
of temperature. Open circles were derived by a cell with 960 atoms,
while filled squares correspond to a cell with 8400 atoms. The lines
are guides to the eye. }
\label{fig:kappa}
\end{figure}

\subsection{Anharmonic effect in $\sigma$,$\;\kappa$, and $\omega_{O}(\Gamma)$\label{sub:anharmonic-effects} }

The linear coefficient, $\sigma,$ of the dispersion relation for
ZA modes in graphene is displayed in Fig. \ref{fig:kappa}a. The results
were obtained from $NPT$ simulations at zero stress and temperatures
between 1 and 2000~K. The coefficient $\sigma$ increases from a vanishing
value in the low temperature (harmonic) limit to a value close to
0.02 eV$\textrm{\AA}^{-2}$ at 2000~K. The increase of $\sigma$ seems
to be linear at low temperatures. Above circa 700~K $\sigma$ grows
less rapidly than linearly. A linear dependence of $\sigma$ with
$T$ was reported in a classical first order perturbation treatment
of graphene as a result of including anharmonic terms in the elastic
model.\citealp{amorim14} The adiabatic treatment of anharmonic effects
of graphene in Ref. \onlinecite{adamyan16} was based on a quantum
description of the out-of-plane vibrations. At temperatures above
700~K they find that the oscillations behave classically and that
$\sigma$ should increase linearly with $T$, or even less than linearly
when the temperature dependence of the in-plane elastic constants
is taken into account. Our simulation results for $\sigma$ are then
in reasonable agreement to the expectations obtained by analytical
treatments of anharmonic effects in the out-of-plane fluctuations
of graphene.\citealp{amorim14,adamyan16}

The temperature dependence of the bending rigidity, $\kappa$, is
displayed in Fig. \ref{fig:kappa}b. The plotted values were derived
from the simulation results via Eq. (\ref{eq:kappa}). Starting from
the harmonic $T\rightarrow0$ limit of the employed LCBOPII model
with $\kappa=1.49$ eV, we observe that $\kappa$ increases linearly
with temperature. Above 700~K the increase becomes slightly slower
than linear. At 2000~K we get a bending rigidity $\kappa\approx2.2$
eV. Previous atomistic simulations of graphene report contradicting
results for the temperature dependence of the bending rigidity. Increase
of $\kappa$ with temperature has been reported in classical Monte
Carlo simulations of graphene.\citealp{costamagna12,zakharchenko10}
However, MD simulations between 200 and 1600~K were reported to present
a decrease in $\kappa$ from 1 eV to 0.4 eV.\citealp{liu09} Even
a temperature independent $\kappa$ has been suggested from MC simulations.\citealp{lajevardipour12}
The determination of $\kappa$ is usually performed by a best fit
of simulated results of $\left\langle \bar{A}_{n}^{2}\right\rangle $
in a $k$-region where the slope can be approximated by the harmonic
behavior of a continuous membrane.\citealp{zakharchenko10,liu09}
However, the atomic character of graphene introduces uncertainty about
the $k-$region where the continuous membrane model is valid. Different
results of $\kappa$ may in part be caused from differences in the
$k$ range where the fit was performed.

\begin{figure}
\vspace{-1.5cm}
\includegraphics[width= 9cm]{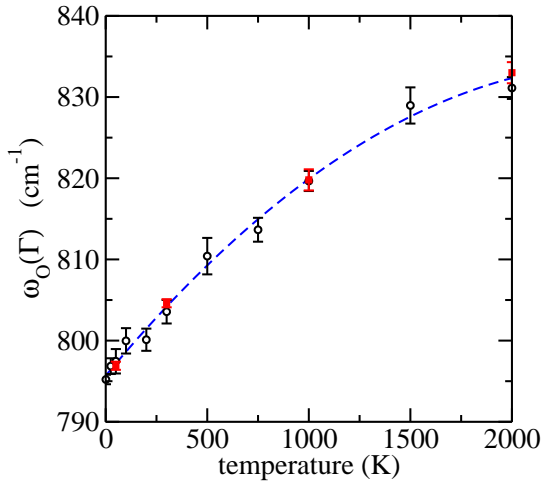}
\vspace{-0.5cm}
\caption{Temperature dependence of the optical ZO mode at the center of the
BZ. Open circles were derived from a cell with 960 atoms, while filled
squares correspond to 8400 atoms. The line is a guide to the eye.}
\label{fig:w_opt}
\end{figure}

The estimated temperature dependence of the ZO mode at the center
of the BZ , $\omega_{O}(\Gamma)$, is displayed in Fig. \ref{fig:w_opt}.
These values were derived from plots of $\omega_{O}$ as a function
of $k$, similar to that presented in Fig. \ref{fig:1K}b at 1~K.
To extrapolate the frequency at $\Gamma$, i.e. at $k=0$, we fitted
a simple relation $\omega_{O}=$$A\cos(Lk)$ to the simulation results
for $k<0.6$ $\textrm{\AA}^{-1}$. The anharmonicity of the employed
model causes an increase of 4\% in the frequency of the optical out-of-plane
mode at $\Gamma$, when the temperature grows up to 2000~K. The linear
increase at low temperatures slows down as the temperature increases,
similarly to the behavior seen before for $\sigma$ and $\kappa$.
This dependence is a consequence of the increase of anharmonic effects
as temperature grows. It is interesting that the effect of temperature
in $\omega_{O}(\Gamma)$ is to make the vibrational mode harder. We
are not aware of any previous prediction about the temperature dependence
of $\omega_{O}(\Gamma)$ in graphene. This increase in vibrational
frequency for rising temperature is similar to that found for acoustic
modes with negative Grüneisen parameter in some solids.\citealp{mounet05}

\subsection{Logarithmic divergence of mean-square heights with sample size\label{sub:divergence_h2} }

The mean-square height fluctuation, $\left\langle h^{2}\right\rangle $,
of the carbon atoms is the sum of the contributions of symmetric,
$\left\langle h_{A}^{2}\right\rangle $ and antisymmetric modes $\left\langle h_{O}^{2}\right\rangle $.
From Eq. (\ref{eq:parseval}), one has

\begin{equation}
\left\langle h_{A}^{2}\right\rangle =\frac{2}{N}\sum_{n=1}^{N_{k}}\left\langle \bar{A}_{n}^{2}\right\rangle \;,
\end{equation}
and a similar relation for $\left\langle h_{O}^{2}\right\rangle $.
We will use an analytical harmonic expression for $\left\langle h_{A}^{2}\right\rangle $
and a simple estimation of $\left\langle h_{O}^{2}\right\rangle $
to rationalize the size dependence of $\left\langle h^{2}\right\rangle $
found in $NPT$ simulations at 300~K with cell sizes up to 33600 atoms. 

For the asymmetric mode we make a rough estimate 

\begin{equation}
\left\langle h_{O}^{2}\right\rangle \approx0.16\:\left\langle h_{A}^{2}\right\rangle \;.\label{eq:h2_O_aprox}
\end{equation}
This relation is derived from our simulation results with 960 and
8400 atoms at 300~K. We find a ratio $\left\langle h_{O}^{2}\right\rangle =0.14\left\langle h_{A}^{2}\right\rangle $
for $N=960$ atoms, while the factor becomes 0.18 for $N=8400$ atoms.
For the sake of simplicity, we have approximated $\left\langle h_{O}^{2}\right\rangle $
for $N<33600$ as the average of both results. 

The analytical harmonic prediction for $\left\langle h_{A}^{2}\right\rangle $
has been derived in Ref. \onlinecite{gao14} under the assumption
that the dispersion relation for the acoustic mode, $\omega_{A}$,
is given by Eq. (\ref{eq: w=00003Dk2_k4}). The parameters $\sigma$
and $\kappa$ of the harmonic model will be taken from our simulation
results with 960 atoms (see Tab. \ref{tab:1}).

\begin{figure}
\vspace{-1.5cm}
\includegraphics[width= 9cm]{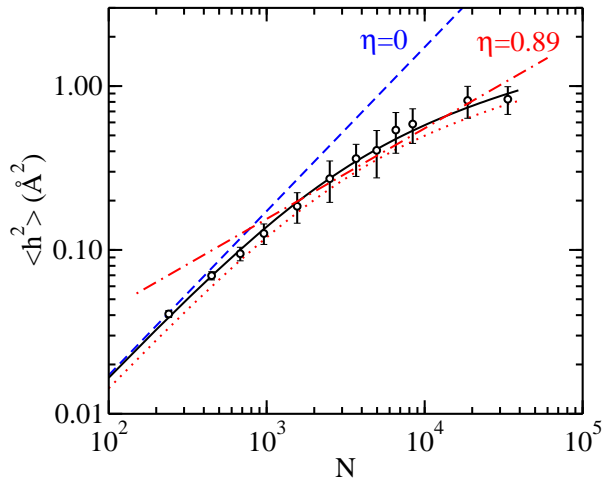}
\vspace{-0.5cm}
\caption{Log-log plot of the mean-square fluctuation of out-of-plane modes
of graphene as a function of the number of atoms in the simulation
cell. Open symbols are results of $NPT$ simulations at zero stress
and $T$=300~K. The dashed line displays the harmonic results for
$\kappa=1.61$ eV and $\sigma=0$, while the continuous line represents
the harmonic limit when $\sigma=0.008$ eV/$\textrm{\AA}^{-1}$. The
curves include contributions from both the symmetric and antisymmetric
modes. The dotted line is derived from the continuous one by subtraction
of the antisymmetric mode contribution. The dash-dotted line is a
power-law fit , $\left\langle h\text{\texttwosuperior}\right\rangle =cN^{1-(\eta/2)}$,
of the simulation results for $N>500$.}
\label{fig:h2}
\end{figure}

Firstly, let us consider the hypothetical case where $\omega_{A}$
has a vanishing linear term (i.e., $\sigma=0$ and therefore $\rho\omega_{A}^{2}=\kappa k^{4}$).
The harmonic limit of the mean-square fluctuations is here\citealp{gao14}

\begin{equation}
\left\langle h_{A1}^{2}\right\rangle =\frac{k_{B}TV_{a}N}{16\pi^{4}\kappa\text{ }}\sum_{j,l=-n}^{n}{}^{'}\left(j^{2}+l^{2}\right)^{-2}\;.\label{eq:h2_A1}
\end{equation}
The sum is over a discrete $k-$point mesh assuming a squared-shaped
membrane with periodic boundary conditions. $n$ is defined as the
number of atoms along each direction of the plane, i.e., $n^{2}=N$.
The prime indicates that the term $j=l=0$ is not included in the
sum. $\left\langle h_{A1}^{2}\right\rangle $ was calculated at 300~K 
as a function of $N$ with the data for $V_{a}$ and $\kappa$ from
Tab. \ref{tab:1}. The total mean-square height is then approximated
under consideration of Eq. (\ref{eq:h2_O_aprox}) as 

\begin{equation}
\left\langle h_{1}^{2}\right\rangle =1.16\left\langle h_{A1}^{2}\right\rangle \;.\label{eq:h2_har1}
\end{equation}
The result as a function of $N$ is displayed in Fig. \ref{fig:h2}
by a dashed line. The expectation values, $\left\langle h^{2}\right\rangle $,
obtained from $NPT$ simulations using Eq. (\ref{eq:h2_def}) for
cell sizes up to 33600 atoms are given by symbols. We note that Eq.
(\ref{eq:h2_har1}) overestimates the out-of-plane height fluctuations
by an unrealistic large amount. 

Secondly, let us include the linear term of our simulations ($\sigma=0.008$
eV$\textrm{\AA}^{-2}$ at 300~K) in the dispersion relation of Eq.
(\ref{eq: w=00003Dk2_k4}). In this case, the harmonic limit of the
mean-square height fluctuations becomes\citealp{gao14}
\begin{multline}
\left\langle h_{A2}^{2}\right\rangle =  \\
\frac{k_{B}TV_{a}N}{16\pi^{4}\kappa\text{ }}\sum_{j,l=-n}^{n}{}^{'}\left(j^{2}+l^{2}\right)^{-2}\left(1+\frac{\sigma V_{a}N}{4\pi^{2}\kappa}\left(j^{2}+l^{2}\right)^{-2}\right)^{-1}\;.\label{eq:h2_A2}
\end{multline}
The result for the total height fluctuation,

\begin{equation}
\left\langle h_{2}^{2}\right\rangle =1.16\left\langle h_{A2}^{2}\right\rangle \;,\label{eq:h2_har2}
\end{equation}
is displayed in Fig. \ref{fig:h2} by a full line. The harmonic model,
with the finite sound velocity corresponding to our value of $\sigma$
at 300~K, gives now a realistic description of the simulation results. 

The dotted line in Fig. \ref{fig:h2} is the symmetric contribution
$\left\langle h_{A2}^{2}\right\rangle $ to the total mean-square
fluctuation. Note that the explicit consideration of the antisymmetric
mode improves the agreement between simulation results of $\left\langle h^{2}\right\rangle $
and the analytical model.

The harmonic approximations, $\left\langle h_{A1}^{2}\right\rangle $
and $\left\langle h_{A2}^{2}\right\rangle $, have different asymptotic
behavior in the limit of large sample size. The summations in Eqs.
(\ref{eq:h2_A1}) and (\ref{eq:h2_A2}) can be converted to integrals
in a continuum approximation. The details are given elsewhere.\citealp{gao14}
Here it suffices to quote that in a continuum limit 

\begin{equation}
\left\langle h_{A1}^{2}\right\rangle \approx\frac{k_{B}TV_{a}N}{16\pi^{3}\kappa\text{ }},
\end{equation}

\begin{equation}
\left\langle h_{A2}^{2}\right\rangle \approx\frac{k_{B}T}{4\pi\sigma\text{ }}\ln\left(1+\frac{\sigma V_{a}N}{4\pi^{2}\kappa\text{ }}\right).
\end{equation}
Thus a finite value of $\sigma$ reduces the divergence of the harmonic
mean-square amplitude, that results proportional to $N$ in $\left\langle h_{A1}^{2}\right\rangle $,
but diverges only logarithmically with $N$ in $\left\langle h_{A2}^{2}\right\rangle $.
The results of Fig. \ref{fig:h2} show that our $NPT$ simulations
at zero stress are in reasonable agreement with a logarithmic divergence
in the long-wavelength behavior of $\left\langle h^{2}\right\rangle $
with the number of atoms $N$. The essential ingredient for this agreement
is the appearance of a linear term, $v_{A}k$, in the dispersion relation
of $\omega_{A}$. This term implies a finite sound velocity for the
long-wavelength limit of the acoustic ZA modes.

\section{Discussion\label{sec:Discussion}}

Best fits presented in the literature of simulated values of mean-square
fluctuations, $\left\langle \bar{A}_{n}^{2}\right\rangle $ or $\left\langle h^{2}\right\rangle $,
should be taken with caution. There is no general agreement about
the theoretically best fitting model. Given that Eq. (\ref{eq:wn_An})
defines a one-to-one correspondence between amplitudes, $\left\langle \bar{A}_{n}^{2}\right\rangle ,$
and frequencies of ZA modes, one can distinguish the models just by
the underlying dispersion relation. 

\begin{table}
\caption{Dispersion relation for ZA modes used in the interpretation of out-of-plane
amplitudes of graphene. $k-$interval gives regions where the model
was applied in simulations. The next column summarizes the large size
limit of the mean-square fluctuations. }
\label{tab:2}%
\begin{tabular}{cccc}
\hline
$\rho\omega_{A}^{2}$ & $k-$interval ($\textrm{\AA}^{-1}$) & $\left\langle h_{A}^{2}\right\rangle $ ($N\rightarrow\infty$) & References\tabularnewline
\hline
$\kappa k^{4}$ & {[}0.3,1{]}\citealp{Los09},{[}0.3,0.4{]}\citealp{Kirilenko2013} & $N$ & \onlinecite{Los09,Kirilenko2013,liu09}\tabularnewline
$\kappa_{r}k^{4-\eta}$ & {[}0,0.2{]}\citealp{Los09},{[}0.4,1{]}\citealp{lajevardipour12} & $N^{1-(\eta/2)}$ & \onlinecite{Los09,gao14,lajevardipour12}\tabularnewline
Eq. (\ref{eq:w_sin_2_DLC}) & $[0,1]$ & $\ln N$ & This work\tabularnewline
\hline
\end{tabular}
\end{table}

Different dispersion relations used for graphene are summarized in
Tab. \ref{tab:2}. Each model reproduces with reasonable accuracy
simulation results in certain $k-$regions. We consider that the apparent
success of fitting simulation data to different models is due to the
fact that information derived from the simulations is always partial.
In particular, the long wavelength limit ($k\rightarrow0)$ is not
easily accessible as the simulation time grows prohibitively with
the cell size ($N\rightarrow\infty)$ and also as the statistics of
very low frequency modes worsens because of limited simulation time.

A further matter of concern is the function to be fitted. Both the
absolute value of the function and the density of $k-$points affect
the result of the least squares method. In the present work we fitted
the function $k_{n}^{2}\left\langle \bar{A}_{n}^{2}\right\rangle $.
The reason is that for a flexural mode with quadratic dispersion the
value of $\left\langle \bar{A}_{n}^{2}\right\rangle ,$ decreases
as $k^{-4}$, while the density of sampled points in $k-$space increases
as $k^{2}$, i.e., as the area of circular sectors of radius $k$.
Thus for $k_{n}^{2}\left\langle \bar{A}_{n}^{2}\right\rangle $, the
value of the fitted function times the density of sampled points becomes
approximately independent of $k$. Evenly distributed weights in $k-$space
is a convenient feature for the least squares method. Let us present
a specific example: if one performs the least squares fit with the
reciprocal function $\left(k_{n}\left\langle \bar{A_{n}}^{2}\right\rangle \right)^{-1},$
instead of $k_{n}^{2}\left\langle \bar{A}_{n}^{2}\right\rangle $,
then the $k_{n}-$points with larger module will effectively have
larger weight in the fit. The reason is that both the value of the
reciprocal function and the density of points increase now as $k^{2}$.
As a consequence, the coefficient $\sigma,$ derived from the best
fit of the reciprocal function, is about two times larger that those
presented in Tab. \ref{tab:1}. 

One important physical difference between the three models in Tab.
\ref{tab:2} is that a finite sound velocity for the acoustic out-of-plane
vibrations, $v_{A}$, is predicted only by the model used in the present
work. The other two dispersion relations in Tab. \ref{tab:2} imply
that $v_{A}=0$ at all temperatures. In this respect, our simulation
results have received an independent confirmation from a recent theoretical
paper that predicts, in terms of a clear physical picture, the acoustic-type
dispersion of the bending mode.\citealp{adamyan16} The origin of
the bending sound velocity is related to the \textit{anharmonic} interaction
between in-plane and out-of-plane vibrations due to non-linear components
in the strain tensor. The investigation by Adamyan \textit{et al.}
shows that the dispersion of the bending mode must be necessarily
linear at small wave numbers.\citealp{adamyan16} We consider this
behavior as an important physical property of graphene that is confirmed
by our analysis of the $NPT$ simulations.

Interestingly, a previous prediction of a linear component in the
dispersion of the ZA mode was presented by Kumar \textit{et al.} \citealp{kumar10}
based on electronic structure density functional theory calculations.
In this work the origin of the rigidity was traced to the coupling
between vibrational and electronic degrees of freedom, arising from
a curvature induced overlap between $\pi$ orbitals in graphene.\citealp{kumar10}
The same result was suggested by Falkovsky\citealp{falkovsky08} by
the study of the symmetry constraints of the phonon dispersion curves
of graphene. The bending velocity could be zero only if a definite
condition is fulfilled for the force constants of the graphene lattice.
The same fact was found for a one-dimensional atomic chain (see Appendix
\ref{appendix:w}). Using the value of force constants obtained by
fitting experimental data of graphite he concluded that graphene possesses
a small but finite bending stiffness.\citealp{falkovsky08} 

Previous classical simulations of out-of-plane fluctuations of graphene
have been analyzed in terms of a power-law behavior of $\left\langle h^{2}\right\rangle $
with a roughness exponent close to one, in agreement to the classical
self-consistent calculation by Nelson and Peliti.\citep{nelson48}
One may wonder what makes our classical simulations to deviate from
this picture. In fact, by fitting our simulated values of $\left\langle h^{2}\right\rangle $
to a power law in Fig. \ref{fig:h2} we get a roughness exponent $\eta=0.89$
(dash-dotted line), close to the values reported in the literature.\citep{Los09,gao14}
We note that the set of cell sizes considered in our simulations is
larger than in previous studies, that used only between 3 and 6 different
cell sizes.\citep{Los09,gao14} The whole range of studied cell sizes
shows that the analytical model for $\left\langle h^{2}\right\rangle $
in Eq. (\ref{eq:h2_A2}) represents an improved agreement to the simulation
results, in comparison to a power-law fit. We stress that the continuous
line in Fig. \ref{fig:h2} is not a fit, but an analytical model defined
with plain physical quantities ($\sigma$ and $\kappa$) derived from
the symmetric out-of-plane fluctuations of our simulations. 

The appearance of a finite bending sound velocity, $v_{A}$, translates
into a roughness exponent $\eta=2$. This is the roughness exponent
obtained by Amorim \textit{et al}.\citep{amorim14} in a quantum self-consistent
perturbative calculation of anharmonic graphene. It may appear surprising
that here the result of our classical simulations agrees with the
roughness exponent of a quantum calculation ($\eta=2$) but disagrees
with the self-consistent classical perturbative result ($\eta=1$).
However, one should consider that the classical simulations include
(numerically) the whole anharmonicity of the employed model potential.
Therefore, the disagreement of the classical simulation with the expectation
of a classical first-order perturbation theory must not be necessarily
considered as a kind of inconsistence. We expect to clarify this issue
in a future work by including quantum effects in our simulations by
the path integral formalism. 

The analysis of the simulated trajectories of graphene, in particular
the study of the asymmetric out-of-plane modes, allowed us the characterization
of the optical bending branch of graphene. This analysis offers additional
physical information that has not been previously considered in simulation
studies of graphene.

\section{Conclusions\label{sec:conclusions}}

We have presented a series of $NPT$ simulations of graphene under
zero stress conditions. The simulations included temperatures up to
2000~K and several cell sizes up to 33600 atoms. The simulations were
performed in the classical limit using the empirical LCBOPII model.
The focus of this study has been the characterization of anharmonic
effects associated to the out-of-plane oscillations of the layer.
The symmetry of the lattice, with two atoms as a basis, imposes the
presence of an acoustic and an optical branch for the out-of-plane
oscillations. We have focused on an atomistic description of both
out-of-plane modes. This description is more general than a continuous
limit of the solid membrane, where optical out-of-plane modes are
absent.

The mean-square out-of-plane fluctuations of carbon atoms have been
analyzed with a model for the dispersion relation of the acoustic
bending branch. The result of this analysis is the characterization
of several anharmonic effects as a function of temperature. The most
important finding is that the mean-square out-of-plane fluctuation
of carbon is compatible with the presence of a linear dispersion term
in the acoustic ZA branch at low wave numbers. This effect is a consequence
of the anharmonicity of the interatomic potential and therefore increases
with temperature as the amplitude of atomic vibrations increases.
The bending sound velocity, derived from the linear dispersion term
of the ZA mode, increases from 0 to 0.6 km/s when the temperature
rises from the zero temperature limit to 2000~K. At the same time
the bending rigidity of graphene is found to increase in this temperature
window from 1.49 eV to 2.2 eV. The frequency of the optical ZO modes
at the $\Gamma$ point of the Brillouin zone displays a shift of 4\%
as temperature increases up to 2000~K. The hardening of these modes
with temperature is again a consequence of the anharmonicity of the
model.

The existence of a finite bending velocity in graphene implies that
the amplitude of the out-of-plane fluctuations, $\left\langle h^{2}\right\rangle $,
is strongly reduced in the long-wavelength limit. If the dispersion
relation of the ZA branch were strictly quadratic, $\left\langle h^{2}\right\rangle $
would diverge proportional to the number of atoms $N$ of the layer.
A finite sound velocity implies that the divergence in $\left\langle h^{2}\right\rangle $
is reduced becoming proportional to $\ln N$. The results of our simulations
are consistent with recent analytical findings.\citealp{adamyan16,amorim14}

\acknowledgments 

This work was supported by Dirección General de Investigación, MINECO
(Spain) through Grants No. FIS2012-31713, FIS2013-47350, and FIS2015-64222-
C2-1-P. The authors benefited from the kind support of J. H. Los in
the implementation of the LCBOPII potential. We thank R. Rold\'an for
a critical reading of the manuscript.

\appendix

\section{Computational conditions\label{appendix:  MD}}

\begin{figure}[!t]
\includegraphics[width= 9cm]{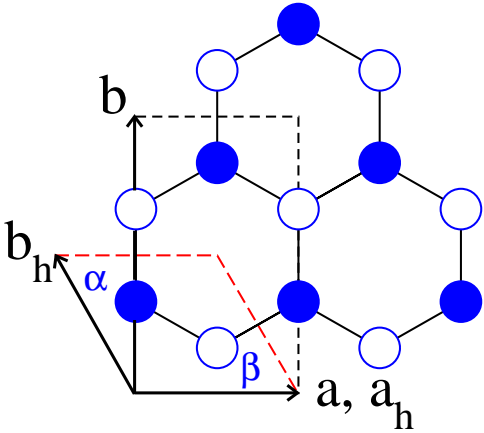}
\vspace{-0.0cm}
\caption{The rectangular cell $(\mathbf{a},\mathbf{b})$ used in the simulations
is displayed together with the standard hexagonal cell $(\mathbf{a}_{h},\mathbf{b}_{h})$
of the graphene lattice. The atomic base of the lattice is made up
of two atoms, labeled as $\alpha$ (full circles) and $\beta$ (open
circles).}
\label{fig:geo}
\end{figure}

The classical MD simulations of graphene were performed on supercells
generated with a two-dimensional (2D) rectangular cell. The relation
of the rectangular axes $(\mathbf{a},\mathbf{b})$ to the standard
hexagonal cell $(\mathbf{a}_{h},\mathbf{b}_{h})$ is shown in Fig.
\ref{fig:geo}. The simulation cell is described by a 2x2 matrix $G$
whose columns are the Cartesian coordinates of the cell vectors 

\begin{equation}
G=(L_{x}\mathbf{a},L_{y}\mathbf{b})\;,
\end{equation}
where $L_{x}$ and $L_{y}$ are positive integers. The supercell $(L_{x},L_{y})$
is chosen to have similar linear dimension in the $x-$and $y-$directions.
We have performed simulation on supercells of several sizes having
between 960 and 33600 atoms. Periodic boundary conditions were applied
to the 2D simulation cell. The area of the simulation cell is $V=|G|$.
It is important to note that the graphene lattice is constructed with
a base of two atoms ($\alpha$ and $\beta$), which are distinguished
in Fig. \ref{fig:geo} as open and closed circles, respectively. In
the ideal lattice each $\beta$ atom is related to an $\alpha$ atom
by a fixed vector

\begin{equation}
\mathbf{u}_{\beta,j}=\mathbf{\mathbf{\mathbf{u}}}_{\alpha,j}+\frac{\mathbf{b}}{3}\;.\label{eq:atoms_AB_app}
\end{equation}

The potential energy $U$ of graphene has been obtained with the LCBOPII
model.\citealp{los03,los05} A slight modification of the original
torsion parameters was made in order to increase the bending constant
of a flat layer in the zero temperature limit from $\kappa=1.1$ eV
to $\kappa=1.48$ eV.\citealp{los16} The latter value is in better
agreement to ab initio electronic structure calculations with values
of $\kappa$ in a range from 1.46 to 1.6 eV.\citealp{politano12}
The temperature was controlled by chains of four Nosé-Hoover thermostats
coupled to each of the Cartesian atomic coordinates. In the case of
the $NPT$ ensemble an additional chain of four barostats was coupled
to the volume.\citep{tu98} To integrate the equations of motion,
a reversible reference system propagator algorithm (RESPA) was employed.\citep{ma96}
For the evolution of thermostats and barostats a time step $\delta t=\Delta t/4$
was used, where $\Delta t$ is the time step associated to the calculation
of forces. A value of $\Delta t$=1 fs was found to provide adequate
convergence, although some check simulations at 1000~K were performed
with a smaller time step of 0.5 fs. Atomic forces were derived analytically
by the derivatives of the potential energy $U$. The stress tensor
estimator was similar to that used in a previous work\citealp{ramirez08c}

\begin{equation}
\sigma_{xy}=\left\langle \frac{1}{V}\left(\sum_{i=1}^{N}mv_{ix}v_{iy}-\frac{\partial U}{\partial\epsilon_{xy}}\right)\right\rangle \;,
\end{equation}
where $v_{ix}$ is a velocity coordinate, $\epsilon_{xy}$ is a component
of the 2D strain tensor, and the brackets $\left\langle \cdots\right\rangle $
indicates an ensemble average. The derivative of $U$ with respect
the strain tensor was performed analytically. Typical runs consisted
of $5\times10^{5}$ MD steps (MDS) for equilibration, followed by
runs using between $2\times10^{6}$ and $8\times10^{6}$ MDS for calculation
of equilibrium properties. Both isotropic and full cell fluctuations
were programmed for the $NPT$ ensemble. The structural analysis was
performed on subsets of $8\times10$\textthreesuperior{} configurations
stored at equidistant times during the whole simulation run. Error
bars were evaluated by dividing the total simulation run into four
blocks and by calculating the standard deviation of the block averages.

\begin{figure}
\includegraphics[width= 9cm]{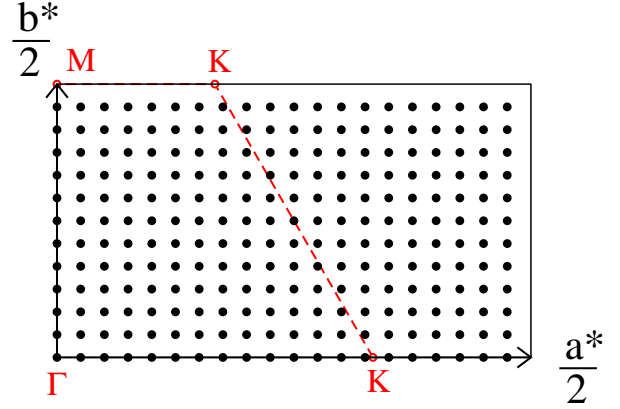}
\vspace{-0.0cm}
\caption{Reciprocal cell $(\mathbf{a^{*}},\mathbf{b^{*}})$ of the rectangular
axes $(\mathbf{a},\mathbf{b})$ employed in the simulations. $\Gamma,M,$
and $K$ label the special positions of the 2D hexagonal BZ. The filled
dots represent the $\mathbf{k_{n}}-$grid used in the Fourier transform
of atom heights for a (20,12) supercell with 960 atoms. }
\label{fig:k}
\end{figure}

The reciprocal lattice corresponding to the simulation cell is defined
by the matrix

\begin{equation}
G_{r}=2\pi\left(G^{-1}\right)^{T}\;.
\end{equation}
The columns of this matrix $(\mathbf{a}^{*},\mathbf{b}^{*})$ define
the wavevectors $\mathbf{k}_{n}$ whose wavelengths are commensurate
with the simulation cell

\begin{equation}
\mathbf{k}_{n}=n_{x}\mathbf{a}^{*}+n_{y}\mathbf{b}^{*}\;,\label{eq:k_n}
\end{equation}
with $n_{x}=0,\ldots,L_{x}-1$, and $n_{y}=0,\ldots,L_{y}-1$. The
total number of $\mathbf{k}_{n}-$points is $N_{k}=L_{x}L_{y}.$ The
$\mathbf{k_{n}}-$grid used in the Fourier transform of heights of
the carbon atoms in a (20,12) supercell is displayed in Fig. \ref{fig:k}. 

\section{Dispersion relation for a linear chain \label{appendix:w}}

The dispersion relation, $\omega(k)$, for a linear chain of atoms
with elastic interactions up to second nearest-neighbors is a a simple
extension of the textbook solution for a first nearest-neighbors interaction.\citealp{kittel}
The interest of this extension is that depending on the relation between
the two force constants ($c_{1}$ and $c_{2})$, one finds three types
of elastic behavior similar to those of a graphene sheet in the harmonic
limit.\citealp{pedro12a,pedro12b} 

Let us consider the atoms along the $x-$axis at coordinates $x_{s}=sL$,
with $L$ being the interatomic distance, and $s$ an integer that
enumerates the atoms. The height of the $s$'th atom with respect
to the axis is $h_{s}$ and the $k-$points are defined in the interval
$\left[-\pi/L,\pi/L\right]$. Considering the collective mode of amplitude
$A$ 

\begin{equation}
h_{s}=Ae^{-ix_{s}k}e^{-i\omega t}\;,\label{eq:h_s}
\end{equation}
the equation of motion of for the $s$'th atom with mass $m$ is

\begin{multline}
m\frac{\partial^{2}h_{s}}{\partial t^{2}}=  \\
c_{1}(h_{s+1}-h_{s})+c_{1}(h_{s-1}-h_{s})+c_{2}(h_{s+2}-h_{s})+c_{2}(h_{s-2}-h_{s})\;.
\end{multline}
Performing the time derivative and taking into account that 
$h_{s+n}=h_{s}e^{-inLk}$, one gets after dividing by $h_{s}$
\begin{equation}
-m\omega^{2}=c_{1}(e^{-iLk}+e^{iLk}-2)+c_{2}(e^{-i2Lk}+e^{i2Lk}-2) \;.
\end{equation}
The sum of the complex exponentials gives a cosine function that is
simplified by the trigonometric relation $1-\cos k=2\sin^{2}(k/2)$
with the result
\begin{equation}
\rho\omega^{2}=D\left[\sin^{2}\left(\frac{Lk}{2}\right)-C\sin^{2}\left(Lk\right)\right]\;.
\end{equation}
with the constants $D=4c_{1}/L$ and $C=-c_{2}/c_{1}$. The atomic
density is $\rho=m/L.$ Depending on the value of $C$ there appears
three different elastic behaviors: ($a$) if $C<$1/4 the dispersion
relation is linear in the long-wavelength limit, $\rho\omega\text{\texttwosuperior}=\sigma k^{2}$
with $\sigma=L(c_{1}+4c_{2})$; ($b$) if $C=$1/4 the dispersion
relation is quadratic for long-wavelength oscillations, $\rho\omega\text{\texttwosuperior}=\kappa k^{4}$,
with $\kappa=c_{1}L^{3}/4$; ($c$) if $C>$1/4 the linear system
is unstable ($\omega\text{\texttwosuperior\ <0 )}$ for $k<(\sigma/\kappa)^{1/2}$.
\bibliographystyle{apsrev}

\begin{thebibliography}{34}
\expandafter\ifx\csname natexlab\endcsname\relax\def\natexlab#1{#1}\fi
\expandafter\ifx\csname bibnamefont\endcsname\relax
  \def\bibnamefont#1{#1}\fi
\expandafter\ifx\csname bibfnamefont\endcsname\relax
  \def\bibfnamefont#1{#1}\fi
\expandafter\ifx\csname citenamefont\endcsname\relax
  \def\citenamefont#1{#1}\fi
\expandafter\ifx\csname url\endcsname\relax
  \def\url#1{\texttt{#1}}\fi
\expandafter\ifx\csname urlprefix\endcsname\relax\def\urlprefix{URL }\fi
\providecommand{\bibinfo}[2]{#2}
\providecommand{\eprint}[2][]{\url{#2}}

\bibitem[{\citenamefont{Meyer et~al.}(2007)\citenamefont{Meyer, Geim,
  Katsnelson, Novoselov, Booth, and Roth}}]{meyer06}
\bibinfo{author}{\bibfnamefont{J.~C.} \bibnamefont{Meyer}},
  \bibinfo{author}{\bibfnamefont{A.~K.} \bibnamefont{Geim}},
  \bibinfo{author}{\bibfnamefont{M.~I.} \bibnamefont{Katsnelson}},
  \bibinfo{author}{\bibfnamefont{K.~S.} \bibnamefont{Novoselov}},
  \bibinfo{author}{\bibfnamefont{T.~J.} \bibnamefont{Booth}}, \bibnamefont{and}
  \bibinfo{author}{\bibfnamefont{S.}~\bibnamefont{Roth}},
  \bibinfo{journal}{Nature} \textbf{\bibinfo{volume}{446}}, \bibinfo{pages}{60}
  (\bibinfo{year}{2007}).

\bibitem[{\citenamefont{Thompson-Flagg
  et~al.}(2009)\citenamefont{Thompson-Flagg, Moura, and Marder}}]{thompson09}
\bibinfo{author}{\bibfnamefont{R.}~\bibnamefont{Thompson-Flagg}},
  \bibinfo{author}{\bibfnamefont{M.}~\bibnamefont{Moura}}, \bibnamefont{and}
  \bibinfo{author}{\bibfnamefont{M.}~\bibnamefont{Marder}},
  \bibinfo{journal}{Europhys. Lett.} \textbf{\bibinfo{volume}{85}},
  \bibinfo{pages}{46002} (\bibinfo{year}{2009}).

\bibitem[{\citenamefont{Amorim et~al.}(2016)\citenamefont{Amorim, Cortijo,
  de~Juan, Grushin, Guinea, Guti\'errez-Rubio, Ochoa, Parente, Rold\'an,
  San-Jose et~al.}}]{amorim16}
\bibinfo{author}{\bibfnamefont{B.}~\bibnamefont{Amorim}},
  \bibinfo{author}{\bibfnamefont{A.}~\bibnamefont{Cortijo}},
  \bibinfo{author}{\bibfnamefont{F.}~\bibnamefont{de~Juan}},
  \bibinfo{author}{\bibfnamefont{A.}~\bibnamefont{Grushin}},
  \bibinfo{author}{\bibfnamefont{F.}~\bibnamefont{Guinea}},
  \bibinfo{author}{\bibfnamefont{A.}~\bibnamefont{Guti\'errez-Rubio}},
  \bibinfo{author}{\bibfnamefont{H.}~\bibnamefont{Ochoa}},
  \bibinfo{author}{\bibfnamefont{V.}~\bibnamefont{Parente}},
  \bibinfo{author}{\bibfnamefont{R.}~\bibnamefont{Rold\'an}},
  \bibinfo{author}{\bibfnamefont{P.}~\bibnamefont{San-Jose}},
  \bibnamefont{et~al.}, \bibinfo{journal}{Phys. Reports}
  \textbf{\bibinfo{volume}{617}}, \bibinfo{pages}{1 } (\bibinfo{year}{2016}).

\bibitem[{\citenamefont{Kirilenko et~al.}(2011)\citenamefont{Kirilenko,
  Dideykin, and Van~Tendeloo}}]{kirilenko11}
\bibinfo{author}{\bibfnamefont{D.~A.} \bibnamefont{Kirilenko}},
  \bibinfo{author}{\bibfnamefont{A.~T.} \bibnamefont{Dideykin}},
  \bibnamefont{and}
  \bibinfo{author}{\bibfnamefont{G.}~\bibnamefont{Van~Tendeloo}},
  \bibinfo{journal}{Phys. Rev. B} \textbf{\bibinfo{volume}{84}},
  \bibinfo{pages}{235417} (\bibinfo{year}{2011}).

\bibitem[{\citenamefont{Gao and Huang}(2014)}]{gao14}
\bibinfo{author}{\bibfnamefont{W.}~\bibnamefont{Gao}} \bibnamefont{and}
  \bibinfo{author}{\bibfnamefont{R.}~\bibnamefont{Huang}}, \bibinfo{journal}{J.
  Mech. Phys. Solids} \textbf{\bibinfo{volume}{66}}, \bibinfo{pages}{42}
  (\bibinfo{year}{2014}).

\bibitem[{\citenamefont{Nelson and Peliti}(1987)}]{nelson48}
\bibinfo{author}{\bibfnamefont{D.}~\bibnamefont{Nelson}} \bibnamefont{and}
  \bibinfo{author}{\bibfnamefont{L.}~\bibnamefont{Peliti}},
  \bibinfo{journal}{J. Phys.} \textbf{\bibinfo{volume}{48}},
  \bibinfo{pages}{1085} (\bibinfo{year}{1987}).

\bibitem[{\citenamefont{Los et~al.}(2009)\citenamefont{Los, Katsnelson, Yazyev,
  Zakharchenko, and Fasolino}}]{Los09}
\bibinfo{author}{\bibfnamefont{J.~H.} \bibnamefont{Los}},
  \bibinfo{author}{\bibfnamefont{M.~I.} \bibnamefont{Katsnelson}},
  \bibinfo{author}{\bibfnamefont{O.~V.} \bibnamefont{Yazyev}},
  \bibinfo{author}{\bibfnamefont{K.~V.} \bibnamefont{Zakharchenko}},
  \bibnamefont{and} \bibinfo{author}{\bibfnamefont{A.}~\bibnamefont{Fasolino}},
  \bibinfo{journal}{Phys. Rev. B} \textbf{\bibinfo{volume}{80}},
  \bibinfo{pages}{121405} (\bibinfo{year}{2009}).

\bibitem[{\citenamefont{Amorim et~al.}(2014)\citenamefont{Amorim, Rold\'an,
  Cappelluti, Fasolino, Guinea, and Katsnelson}}]{amorim14}
\bibinfo{author}{\bibfnamefont{B.}~\bibnamefont{Amorim}},
  \bibinfo{author}{\bibfnamefont{R.}~\bibnamefont{Rold\'an}},
  \bibinfo{author}{\bibfnamefont{E.}~\bibnamefont{Cappelluti}},
  \bibinfo{author}{\bibfnamefont{A.}~\bibnamefont{Fasolino}},
  \bibinfo{author}{\bibfnamefont{F.}~\bibnamefont{Guinea}}, \bibnamefont{and}
  \bibinfo{author}{\bibfnamefont{M.~I.} \bibnamefont{Katsnelson}},
  \bibinfo{journal}{Phys. Rev. B} \textbf{\bibinfo{volume}{89}},
  \bibinfo{pages}{224307} (\bibinfo{year}{2014}).

\bibitem[{\citenamefont{Adamyan et~al.}(2015)\citenamefont{Adamyan, Bondarev,
  and Zavalniuk}}]{adamyan16}
\bibinfo{author}{\bibfnamefont{V.~M.} \bibnamefont{Adamyan}},
  \bibinfo{author}{\bibfnamefont{V.~N.} \bibnamefont{Bondarev}},
  \bibnamefont{and} \bibinfo{author}{\bibfnamefont{V.~V.}
  \bibnamefont{Zavalniuk}}, \emph{\bibinfo{title}{Bending sound in graphene:
  origin and manifestation}} (\bibinfo{publisher}{arXiv:1510.07878},
  \bibinfo{year}{2015}).

\bibitem[{\citenamefont{Los and Fasolino}(2003)}]{los03}
\bibinfo{author}{\bibfnamefont{J.~H.} \bibnamefont{Los}} \bibnamefont{and}
  \bibinfo{author}{\bibfnamefont{A.}~\bibnamefont{Fasolino}},
  \bibinfo{journal}{Phys. Rev. B} \textbf{\bibinfo{volume}{68}},
  \bibinfo{pages}{024107} (\bibinfo{year}{2003}).

\bibitem[{\citenamefont{Los et~al.}(2005)\citenamefont{Los, Ghiringhelli,
  Meijer, and Fasolino}}]{los05}
\bibinfo{author}{\bibfnamefont{J.~H.} \bibnamefont{Los}},
  \bibinfo{author}{\bibfnamefont{L.~M.} \bibnamefont{Ghiringhelli}},
  \bibinfo{author}{\bibfnamefont{E.~J.} \bibnamefont{Meijer}},
  \bibnamefont{and} \bibinfo{author}{\bibfnamefont{A.}~\bibnamefont{Fasolino}},
  \bibinfo{journal}{Phys. Rev. B} \textbf{\bibinfo{volume}{72}},
  \bibinfo{pages}{214102} (\bibinfo{year}{2005}).

\bibitem[{\citenamefont{Ashcroft and Mermin}(1976)}]{ashcroft}
\bibinfo{author}{\bibfnamefont{N.~W.} \bibnamefont{Ashcroft}} \bibnamefont{and}
  \bibinfo{author}{\bibfnamefont{D.~N.} \bibnamefont{Mermin}},
  \emph{\bibinfo{title}{{Solid State Physics}}} (\bibinfo{publisher}{Saunders
  College}, \bibinfo{address}{Philadelphia}, \bibinfo{year}{1976}).

\bibitem[{\citenamefont{Ram\'irez and L\'opez-Ciudad}(2001)}]{ramirez01}
\bibinfo{author}{\bibfnamefont{R.}~\bibnamefont{Ram\'irez}} \bibnamefont{and}
  \bibinfo{author}{\bibfnamefont{T.}~\bibnamefont{L\'opez-Ciudad}},
  \bibinfo{journal}{J. Chem. Phys.} \textbf{\bibinfo{volume}{115}},
  \bibinfo{pages}{103} (\bibinfo{year}{2001}).

\bibitem[{\citenamefont{Ram\'irez and Herrero}(2005)}]{ramirez05}
\bibinfo{author}{\bibfnamefont{R.}~\bibnamefont{Ram\'irez}} \bibnamefont{and}
  \bibinfo{author}{\bibfnamefont{C.~P.} \bibnamefont{Herrero}},
  \bibinfo{journal}{Phys. Rev. B} \textbf{\bibinfo{volume}{72}},
  \bibinfo{pages}{024303} (\bibinfo{year}{2005}).

\bibitem[{\citenamefont{Herrero and Ram\'{\i}rez}(2009)}]{herrero09}
\bibinfo{author}{\bibfnamefont{C.~P.} \bibnamefont{Herrero}} \bibnamefont{and}
  \bibinfo{author}{\bibfnamefont{R.}~\bibnamefont{Ram\'{\i}rez}},
  \bibinfo{journal}{Phys. Rev. B} \textbf{\bibinfo{volume}{80}},
  \bibinfo{pages}{035207} (\bibinfo{year}{2009}).

\bibitem[{\citenamefont{Herrero and Ram\'{\i}rez}(2010)}]{herrero10}
\bibinfo{author}{\bibfnamefont{C.~P.} \bibnamefont{Herrero}} \bibnamefont{and}
  \bibinfo{author}{\bibfnamefont{R.}~\bibnamefont{Ram\'{\i}rez}},
  \bibinfo{journal}{Phys. Rev. B} \textbf{\bibinfo{volume}{82}},
  \bibinfo{pages}{174117} (\bibinfo{year}{2010}).

\bibitem[{\citenamefont{Lambin}(2014)}]{lambin14}
\bibinfo{author}{\bibfnamefont{P.}~\bibnamefont{Lambin}},
  \bibinfo{journal}{Appl. Sci.} \textbf{\bibinfo{volume}{4}},
  \bibinfo{pages}{282} (\bibinfo{year}{2014}).

\bibitem[{\citenamefont{Rold\'an et~al.}(2011)\citenamefont{Rold\'an, Fasolino,
  Zakharchenko, and Katsnelson}}]{roldan11}
\bibinfo{author}{\bibfnamefont{R.}~\bibnamefont{Rold\'an}},
  \bibinfo{author}{\bibfnamefont{A.}~\bibnamefont{Fasolino}},
  \bibinfo{author}{\bibfnamefont{K.~V.} \bibnamefont{Zakharchenko}},
  \bibnamefont{and} \bibinfo{author}{\bibfnamefont{M.~I.}
  \bibnamefont{Katsnelson}}, \bibinfo{journal}{Phys. Rev. B}
  \textbf{\bibinfo{volume}{83}}, \bibinfo{pages}{174104}
  (\bibinfo{year}{2011}).

\bibitem[{\citenamefont{Costamagna et~al.}(2012)\citenamefont{Costamagna,
  Neek-Amal, Los, and Peeters}}]{costamagna12}
\bibinfo{author}{\bibfnamefont{S.}~\bibnamefont{Costamagna}},
  \bibinfo{author}{\bibfnamefont{M.}~\bibnamefont{Neek-Amal}},
  \bibinfo{author}{\bibfnamefont{J.~H.} \bibnamefont{Los}}, \bibnamefont{and}
  \bibinfo{author}{\bibfnamefont{F.~M.} \bibnamefont{Peeters}},
  \bibinfo{journal}{Phys. Rev. B} \textbf{\bibinfo{volume}{86}},
  \bibinfo{pages}{041408} (\bibinfo{year}{2012}).

\bibitem[{\citenamefont{Zakharchenko et~al.}(2010)\citenamefont{Zakharchenko,
  Los, Katsnelson, and Fasolino}}]{zakharchenko10}
\bibinfo{author}{\bibfnamefont{K.~V.} \bibnamefont{Zakharchenko}},
  \bibinfo{author}{\bibfnamefont{J.~H.} \bibnamefont{Los}},
  \bibinfo{author}{\bibfnamefont{M.~I.} \bibnamefont{Katsnelson}},
  \bibnamefont{and} \bibinfo{author}{\bibfnamefont{A.}~\bibnamefont{Fasolino}},
  \bibinfo{journal}{Phys. Rev. B} \textbf{\bibinfo{volume}{81}},
  \bibinfo{pages}{235439} (\bibinfo{year}{2010}).

\bibitem[{\citenamefont{Liu and Zhang}(2009)}]{liu09}
\bibinfo{author}{\bibfnamefont{P.}~\bibnamefont{Liu}} \bibnamefont{and}
  \bibinfo{author}{\bibfnamefont{Y.~W.} \bibnamefont{Zhang}},
  \bibinfo{journal}{Appl. Phys. Lett.} \textbf{\bibinfo{volume}{94}},
  \bibinfo{eid}{231912} (\bibinfo{year}{2009}).

\bibitem[{\citenamefont{Lajevardipour et~al.}(2012)\citenamefont{Lajevardipour,
  Neek-Amal, and Peeters}}]{lajevardipour12}
\bibinfo{author}{\bibfnamefont{A.}~\bibnamefont{Lajevardipour}},
  \bibinfo{author}{\bibfnamefont{M.}~\bibnamefont{Neek-Amal}},
  \bibnamefont{and} \bibinfo{author}{\bibfnamefont{F.~M.}
  \bibnamefont{Peeters}}, \bibinfo{journal}{J. Phys.: Condens. Matter}
  \textbf{\bibinfo{volume}{24}}, \bibinfo{pages}{175303}
  (\bibinfo{year}{2012}).

\bibitem[{\citenamefont{Mounet and Marzari}(2005)}]{mounet05}
\bibinfo{author}{\bibfnamefont{N.}~\bibnamefont{Mounet}} \bibnamefont{and}
  \bibinfo{author}{\bibfnamefont{N.}~\bibnamefont{Marzari}},
  \bibinfo{journal}{Phys. Rev. B} \textbf{\bibinfo{volume}{71}},
  \bibinfo{pages}{205214} (\bibinfo{year}{2005}).

\bibitem[{\citenamefont{Kumar et~al.}(2010)\citenamefont{Kumar, Hembram, and
  Waghmare}}]{kumar10}
\bibinfo{author}{\bibfnamefont{S.}~\bibnamefont{Kumar}},
  \bibinfo{author}{\bibfnamefont{K.~P. S.~S.} \bibnamefont{Hembram}},
  \bibnamefont{and} \bibinfo{author}{\bibfnamefont{U.~V.}
  \bibnamefont{Waghmare}}, \bibinfo{journal}{Phys. Rev. B}
  \textbf{\bibinfo{volume}{82}}, \bibinfo{pages}{115411}
  (\bibinfo{year}{2010}).

\bibitem[{\citenamefont{Falkovsky}(2008)}]{falkovsky08}
\bibinfo{author}{\bibfnamefont{L.}~\bibnamefont{Falkovsky}},
  \bibinfo{journal}{Phys. Lett. A} \textbf{\bibinfo{volume}{372}},
  \bibinfo{pages}{5189 } (\bibinfo{year}{2008}).

\bibitem[{\citenamefont{Los}(2016)}]{los16}
\bibinfo{author}{\bibfnamefont{J.~H.} \bibnamefont{Los}}
  (\bibinfo{year}{2016}), \bibinfo{note}{private communication}.

\bibitem[{\citenamefont{Politano et~al.}(2012)\citenamefont{Politano, Marino,
  Campi, Far\'ias, Miranda, and Chiarello}}]{politano12}
\bibinfo{author}{\bibfnamefont{A.}~\bibnamefont{Politano}},
  \bibinfo{author}{\bibfnamefont{A.~R.} \bibnamefont{Marino}},
  \bibinfo{author}{\bibfnamefont{D.}~\bibnamefont{Campi}},
  \bibinfo{author}{\bibfnamefont{D.}~\bibnamefont{Far\'ias}},
  \bibinfo{author}{\bibfnamefont{R.}~\bibnamefont{Miranda}}, \bibnamefont{and}
  \bibinfo{author}{\bibfnamefont{G.}~\bibnamefont{Chiarello}},
  \bibinfo{journal}{Carbon} \textbf{\bibinfo{volume}{50}}, \bibinfo{pages}{4903
  } (\bibinfo{year}{2012}).

\bibitem[{\citenamefont{Tuckerman and Hughes}(1998)}]{tu98}
\bibinfo{author}{\bibfnamefont{M.~E.} \bibnamefont{Tuckerman}}
  \bibnamefont{and} \bibinfo{author}{\bibfnamefont{A.}~\bibnamefont{Hughes}},
  in \emph{\bibinfo{booktitle}{Classical \& Quantum Dynamics in Condensed Phase
  Simulations}}, edited by \bibinfo{editor}{\bibfnamefont{B.~J.}
  \bibnamefont{Berne}} \bibnamefont{and} \bibinfo{editor}{\bibfnamefont{D.~F.}
  \bibnamefont{Coker}} (\bibinfo{publisher}{Word Scientific},
  \bibinfo{address}{Singapore}, \bibinfo{year}{1998}), p. \bibinfo{pages}{311}.

\bibitem[{\citenamefont{Martyna et~al.}(1996)\citenamefont{Martyna, Tuckerman,
  Tobias, and Klein}}]{ma96}
\bibinfo{author}{\bibfnamefont{G.~J.} \bibnamefont{Martyna}},
  \bibinfo{author}{\bibfnamefont{M.~E.} \bibnamefont{Tuckerman}},
  \bibinfo{author}{\bibfnamefont{D.~J.} \bibnamefont{Tobias}},
  \bibnamefont{and} \bibinfo{author}{\bibfnamefont{M.~L.} \bibnamefont{Klein}},
  \bibinfo{journal}{Mol. Phys.} \textbf{\bibinfo{volume}{87}},
  \bibinfo{pages}{1117} (\bibinfo{year}{1996}).

\bibitem[{\citenamefont{Ram\'\i{}rez et~al.}(2008)\citenamefont{Ram\'\i{}rez,
  Herrero, Hern\'andez, and Cardona}}]{ramirez08c}
\bibinfo{author}{\bibfnamefont{R.}~\bibnamefont{Ram\'\i{}rez}},
  \bibinfo{author}{\bibfnamefont{C.~P.} \bibnamefont{Herrero}},
  \bibinfo{author}{\bibfnamefont{E.~R.} \bibnamefont{Hern\'andez}},
  \bibnamefont{and} \bibinfo{author}{\bibfnamefont{M.}~\bibnamefont{Cardona}},
  \bibinfo{journal}{Phys. Rev. B} \textbf{\bibinfo{volume}{77}},
  \bibinfo{pages}{045210} (\bibinfo{year}{2008}).

\bibitem[{\citenamefont{Kittel}(1966)}]{kittel}
\bibinfo{author}{\bibfnamefont{C.}~\bibnamefont{Kittel}},
  \emph{\bibinfo{title}{Introduction to {S}olid {S}tate {P}hysics}}
  (\bibinfo{publisher}{Wiley, New York}, \bibinfo{year}{1966}).

\bibitem[{\citenamefont{de~Andres
  et~al.}(2012{\natexlab{a}})\citenamefont{de~Andres, Guinea, and
  Katsnelson}}]{pedro12a}
\bibinfo{author}{\bibfnamefont{P.~L.} \bibnamefont{de~Andres}},
  \bibinfo{author}{\bibfnamefont{F.}~\bibnamefont{Guinea}}, \bibnamefont{and}
  \bibinfo{author}{\bibfnamefont{M.~I.} \bibnamefont{Katsnelson}},
  \bibinfo{journal}{Phys. Rev. B} \textbf{\bibinfo{volume}{86}},
  \bibinfo{pages}{144103} (\bibinfo{year}{2012}{\natexlab{a}}).

\bibitem[{\citenamefont{de~Andres
  et~al.}(2012{\natexlab{b}})\citenamefont{de~Andres, Guinea, and
  Katsnelson}}]{pedro12b}
\bibinfo{author}{\bibfnamefont{P.~L.} \bibnamefont{de~Andres}},
  \bibinfo{author}{\bibfnamefont{F.}~\bibnamefont{Guinea}}, \bibnamefont{and}
  \bibinfo{author}{\bibfnamefont{M.~I.} \bibnamefont{Katsnelson}},
  \bibinfo{journal}{Phys. Rev. B} \textbf{\bibinfo{volume}{86}},
  \bibinfo{pages}{245409} (\bibinfo{year}{2012}{\natexlab{b}}).

\bibitem[{\citenamefont{Kirilenko}(2013)}]{Kirilenko2013}
\bibinfo{author}{\bibfnamefont{D.~A.} \bibnamefont{Kirilenko}},
  \bibinfo{journal}{Tech. Phys. Lett.} \textbf{\bibinfo{volume}{39}},
  \bibinfo{pages}{325} (\bibinfo{year}{2013}).

\end{thebibliography}

\end{document}